\documentstyle[aps,pre,eqsecnum,multicol]{revtex}

\begin{document}

\draft

\title{Spectra of turbulence in dilute
polymer solutions}

\author{A.~Fouxon$^a$ and V.~Lebedev$^{b,c}$}

\address{$^a$ Physics Department, Weizmann Institute of Science,
Rehovot, 76100, Israel \\
$^b$ Landau Inst. for Theoretical Physics, Moscow,
Kosygina 2, 117940, Russia \\
$^c$ Theoretical Division, LANL, Los Alamos, NM 87545, USA}

\date{\today}

\maketitle

\begin{abstract}

  We investigate turbulence in dilute polymer solutions when polymers
  are strongly stretched by the flow. We establish power-law spectrum
  of velocity, which is not associated with a flux of a conserved quantity,
  in two cases. The first case is the elastic waves range of high Reynolds
  number turbulence of polymer solutions above the coil-stretch transition.
  The second case is the elastic turbulence, where chaotic flow is excited
  due to elastic instabilities at small Reynolds numbers.

\end{abstract}

\pacs{PACS numbers 83.50.Ws, 61.25.Hq, 47.27.-i, 05.40.-a}

\begin{multicols}{2}

\section*{Introduction}

In this paper we continue theoretical investigation of turbulence
in dilute polymer solutions, started in \cite{short,01BFL}. As
opposed to Newtonian fluids, such solutions possess additional
macroscopic degrees of freedom related to the elasticity of the
polymer molecules. Relaxation times of elastic stresses can be
comparable with time-scales of the flow which means that the
relation between the stress and the velocity gradient is non-local
in time and, consequently, in space. It is a striking property of
dilute polymer solutions that minute amounts of polymer can
significantly modify the flow. Probably, the most famous example is
the drag reduction phenomenon. Addition of long-chain polymers in
concentrations as small as, say, $10^{-5}$ by weight, can induce a
substantial reduction of the drag force needed to push a turbulent
fluid through a pipe \cite{Virk,McComb,95GB}. Another example is
the elastic turbulence \cite{00GS,01GSa,01GSb}, which is a chaotic
flow, excited in the dilute polymer solutions at low Reynolds numbers.

The reason why small amounts of polymer can significantly modify
properties of the fluid is flexibility of polymer molecules. At
equilibrium a polymer molecule coils up into a spongy ball of a
radius $R_0$. The value of $R_0$ depends on the number of monomers
in the molecule, which is usually very large. For a dilute solution
with the concentration of the molecules, $n$, satisfying
$nR_0^3\ll1$, an influence of equilibrium size molecules on the
hydrodynamic properties of the fluid can be neglected. When placed
in an inhomogeneous flow, such a molecule is deformed into an
elongated structure, which can be characterized by its end-to-end
distance $R$. If the number of monomers in a typical polymer
molecule is large, the elongation $R$ can be much larger than
$R_0$. The influence of the molecules on the flow increases with
their elongation and may become substantial when $R\gg R_0$.

Deformation of a polymer molecule is determined by two processes,
stretching by the velocity gradients and relaxation due to
elasticity of the molecule. To understand how the molecule resists
the deformation by the flow, one should consider its relaxation.
Recent experiments with DNA molecules indicate that the relaxation
is linear in the wide region of scales $R_0\ll R\ll R_{\rm max}$,
where $R_{\rm max}$ is the maximum molecule extension \cite{Chu1}.
In the case of polymers, theoretical arguments and numerics
presented in \cite{99HQ} support the linear relaxation. These
results can be understood if we assume that at $R\gg R_0$ the role
of excluded volume and hydrodynamic interactions between the
monomers are negligible. Then the random walk arguments suggest
that the entropy of polymer molecules is quadratic in $R$ in the
range $R_0\ll R\ll R_{\rm max}$ implying linear relaxation. Whether
the polymers are excited by the flow is determined by the softest
relaxation mode that corresponds to the dynamics of the elongation
$R$. In the absence of stretching, the relaxation of $R$ is
described by the equation $\partial_t{R}=-R/\tau$, where $\tau$ is
a relaxation time, which is expected to be $R$-independent at
$R_0\ll R\ll R_{\rm max}$. If the end-to-end distance $R$ is of the
order of the maximum extension, $\tau$ starts to depend on $R$ and
the dynamics of the molecule becomes nonlinear \cite{87BCAH}.
Possible statistical consequences of the non-linearity have been
investigated in \cite{misha}.

The behavior of the molecule in an inhomogeneous steady flow
depends on the value of the Weissenberg number, ${\rm Wi}$, defined
as the product of the characteristic velocity gradient and $\tau$.
When a polymer molecule is placed in a flow, smooth at the scale
$R$, the velocity difference between the end-points is proportional
to $R$ multiplied by the characteristic value of the velocity
gradient. At $\mbox{Wi}\ll1$ the relaxation is fast as compared to
the stretching time and the polymer always relaxes to the
equilibrium size, $R_0$. The behavior of the polymer at ${\rm
Wi}\gtrsim 1$ depends on the geometry of the flow. For purely
elongational flows the molecule gets aligned along the principal
stretching direction. If the velocity gradient is larger than the
inverse relaxation time, i.e. ${\rm Wi}\gtrsim 1$, the elastic
response becomes too slow in comparison with the stretching and the
molecule gets substantially elongated \cite{Lumley73}. The sharp
transition from the coiled state to the strongly extended state is
called the coil-stretch transition. Rotation can suppress the
transition and even damp it completely since the molecule does not
always point in the stretching direction (see, e.g.,
\cite{Lumley72}). For example, no coil-stretch transition occurs in
the case of the shear flow, which is a particular combination of
the elongational and rotational flows.

In contrast to the steady case, a polymer molecule, moving in a
random flow, alternately enters regions of high and low stretching.
As the intensity of the flow increases, the effect of the
stretching becomes more pronounced. One can generally assert
existence of the coil-stretch transition in this case. This
has been first demonstrated by Lumley \cite{Lumley72}, who
considered the situation where the characteristic time of
variations of the velocity gradient is much larger than the inverse
of the characteristic value of the gradient. He showed that if the
amplitude of the velocity gradient fluctuations is large enough,
the expectation value of $R^2$ grows with time, which signifies the
coil-stretch transition. We have shown in \cite{short,01BFL} that
the coil-stretch transition occurs in any random flow and
established a general criterion for the transition. In particular,
the transition occurs in the situation where the time of velocity
gradient variation is of the order of the inverse of its
characteristic value, which is likely to be the case for real
flows. The coil-stretch transition in random flows is controlled by
the parameter $\lambda_1\tau$, where $\lambda_1$ is the average
logarithmic divergence rate of nearby Lagrangian trajectories, to
be referred to as the principal Lyapunov exponent (which is
positive for an incompressible flow \cite{Furst,Zeldovich}). The
molecules are weakly stretched if $\lambda_1\tau<1$ and strongly
stretched otherwise. Therefore for random flows the parameter
$\lambda_1\tau$ plays the role of the Weissenberg number.

As it is well known (see, e.g., \cite{Frisch,MY}), turbulent flows in
Newtonian fluids consist of chaotic eddies from a wide interval of
scales, $\eta< r< L$, where $L$ is the integral scale (where the
flow is excited) and $\eta$ is the viscous scale. The energy pumped
at the scale $L$ cascades down to the scale $\eta$, where it
dissipates. The size of the polymer molecules is usually much
smaller than the viscous scale. Viscosity makes the flow smooth at
scales $r<\eta$, i.e. the velocity can be approximated by linear
profiles at these scales. Therefore, if $R<\eta$, then the
stretching of molecules is determined by the velocity gradient,
which is random in the turbulent flow. The Lyapunov exponent can be
estimated as the characteristic value of the velocity gradient,
which is determined by the eddies at the viscous scale $\eta$. As
the Reynolds number grows, the velocity gradient increases, and
so does $\lambda_1\tau$. At some value $\mbox{Re}_c$ of the
Reynolds number the product $\lambda_1\tau$ reaches the value $1$
and the coil-stretch transition occurs.

Several mechanisms can limit the polymer stretching above the
coil-stretch transition. The first one is the internal
non-linearity of the elasticity of the polymer molecules. If this
mechanism dominates, then above the transition the molecules are
stretched up to the elongation of the order of $R_{\rm max}$. An
alternative mechanism has been proposed by Tabor and de Gennes
\cite{Tabor}. It is based on the assumption that the elongation of
the polymer molecules $R$ becomes larger than the viscous length of
turbulence, $\eta$, where the elastic force always wins over the
stretching at a certain value of the elongation. Below, we assume
that $R\ll\eta$, which seems to be reasonable for typical polymer
solutions. Another limiting mechanism is the back reaction of the
polymers on the flow. It is caused by the collective contribution
of coherently deformed polymer molecules into the stress tensor.
This elastic part of the stress grows with the molecule elongation.
When it becomes of the order of the viscous stresses existing in
the flow, the polymers modify the flow around them suppressing the
stretching. As a result, a dynamic equilibrium is realized at a
characteristic elongation, $R_{\rm back}$. The total polymer stress
is proportional to $nR^2$, so that $R_{\rm back}$ depends on the
polymer concentration $n$. We assume that the concentration is
large enough for the value of $R_{\rm back}$ to be much smaller
than $R_{\rm max}$. Probably, the condition $R_{\rm back}\ll R_{\rm
max}$ is necessary for existence of a stationary state, because the
polymer molecules, stretched up to $R_{\rm max}$, are intensively
destroyed by the flow.

Above the coil-stretch transition the back reaction modifies the
small-scale properties of turbulent flows, which leads to the
emergence of a new scale, $r_*>\eta$, where energy dissipates
mainly due to polymer relaxation. The scale $r_*$ plays the role of
a new dissipation scale. Large-scale eddies with the sizes $r>r_*$
do not excite elastic degrees of freedom so the usual inertial
energy cascade is realized at these scales. At $\mbox{Re}\gg
\mbox{Re}_c$ there appears a new region of scales, $\eta_*<r<r_*$,
where elastic waves can propagate \cite{01BFL}, which are analogous
to the Alfven waves in magnetic hydrodynamics. At the scale
$\eta_*$ viscosity becomes essential, leading to the strong damping
of the elastic waves.

In this work we investigate the velocity spectrum in the elastic
wave range $k r_*\gg 1$ (where $\bbox k$ is the wave vector). We
show that the spectrum obeys a power law. The ideas of the analysis
go back to the works of Townsend \cite{Townsend} and Batchelor
\cite{Batchelor'}. They recognized that fluctuations with scales,
smaller than the smoothness scale of the flow ($\eta$ in the case
of usual turbulence and $r_*$ in our case), evolve in the linear
velocity profiles. For the passive scalar at large Prandtl numbers
Batchelor derived the spectrum with the power-law $k^{-1}$
\cite{Batchelor'}, originating from the exponential character of
stretching in the linear flow. Formally, it is explained by zero
scaling dimension of the advection term that implies scale
invariance. This property is not broken by a linear decay term
where the power-law spectrum still holds, as it was shown for a
linearly decaying passive scalar by Chertkov \cite{98Che}. Here we
introduce a consistent theoretical scheme for the description of
the small-scale fluctuations and show that though the dynamics of
the small-scale fluctuations is more complicated, than in the case
of the passive scalar (advection and linear decay accompanied by
stretching and waves), the power-law in the spectrum still holds.
The wave oscillations break the scale-invariance, but their
influence on the energy balance is reduced to forcing the
equipartition of kinetic and elastic energies of small-scale
fluctuations. The power law spectrum terminates at $k\sim
(\nu\tau)^{-1/2}$, where viscosity overcomes stretching. The
power-law interval widens as $\mbox{Re}$ grows. Let us stress that
this law is not related to a flux of any conserved quantity.

Another situation, where a power-law spectrum of the small-scale
fluctuations is observed, is the elastic turbulence, realized in
low-Reynolds polymer solutions, if the Weissenberg number
$\mbox{Wi}$ is large enough \cite{00GS,01GSa,01GSb}. It was shown
experimentally in \cite{00GS,01GSa,01GSb}, that the coil-stretch
transition, occurring at increasing $\mbox{Wi}$, leads to a chaotic
flow even though $\mbox{Re}$ is small. Its existence is due to
hydrodynamic instabilities caused by the presence of the elastic
stresses. The velocity spectrum is observed to be power-like in a
wide range of scales in this case \cite{00GS,01GSa,01GSb}. We
demonstrate that, in contrast to the usual hydrodynamic turbulence,
in the elastic turbulence the power velocity spectrum is not
related to the energy cascade, since the main energy dissipation
occurs at the largest scales. The mechanism, leading to this power
spectrum is, again, similar to the linearly decaying passive scalar
problem.

The structure of the paper is as follows. In Section \ref{basiceq}
we introduce a system of equations describing the coupled dynamics
of inertial and elastic degrees of freedom. This system is similar
to the system of equations describing the magnetohydrodynamics
(MHD) \cite{MHD} with the important difference of a linear decay
term in the equation on the ``magnetic field''. In Section
\ref{sec:passive} we present results, concerning statistics of a
passive scalar with a constant damping, embedded in a random flow.
It is a prototype for the subsequent consideration. Section
\ref{sec:saturation} is devoted to the description of the principal
properties of the large $\mbox{Re}$ turbulence in the presence of
polymers and to the derivation of the power-law spectrum in the
elastic dissipation range. In Section \ref{sec:elastic} we
establish the power-law spectrum for the elastic turbulence. In
Conclusion we summarize our results and discuss possible
implications of our work for other subjects. Appendix is devoted to
some details of the Lagrangian statistics.

\section{Basic Relations}
\label{basiceq}

We study dynamics of dilute polymer solutions at scales much larger
than the inter-molecular distance where the polymer solution can be
regarded as a continuous medium and described by macroscopic
fields. Characteristic times of considered processes are regarded
to be comparable with the polymer relaxation time $\tau$. In this
case, besides the usual hydrodynamic degrees of freedom, one has to
take into account degrees of freedom, related to the polymer
elasticity. These degrees of freedom can be described in terms of
the polymer stress tensor \cite{87BCAH}.

We assume that the flow can be treated as incompressible, that is
$\nabla\cdot\bbox v=0$, where $\bbox v$ is the velocity of the
flow. This is justified provided processes at a given scale are
slow in comparison with sound oscillations at the same scale. Then
the velocity dynamics can be described in terms of the following
equation \cite{87BCAH}
\begin{eqnarray} &&
\partial_t v_i+(\bbox v\nabla)v_i
+\varrho^{-1}\nabla_i P=\nu \nabla^2 v_i
+\nabla_j \Pi_{ij} \,,
\label{intro1} \end{eqnarray}
which is a generalization of the Navier-Stokes equation to the case
of viscoelastic fluids. Here $P$ is the pressure, $\nu$ is the
kinematic viscosity of the solvent, $\varrho$ is the fluid mass
density, and $\Pi_{ij}$ is the polymer contribution into the
stress tensor per unit mass.

The equation (\ref{intro1}) has to be supplemented by an equation
for the polymer stress tensor $\Pi_{ij}$ \cite{87BCAH}. We assume
the following equation
\begin{eqnarray} &&
\partial_t {\Pi}_{ij}+(\bbox v\nabla){\Pi}_{ij}
={\Pi}_{kj}\nabla_k v_i+{\Pi}_{ik}\nabla_k v_j
\nonumber \\ &&
-\frac{2}{\tau}\left({\Pi}_{ij}-\Pi_0\delta_{ij}\right) \,,
\label{intro5} \end{eqnarray}
where $\tau$ is the polymer relaxation time, and $\Pi_0$ is related
to the thermal fluctuations of the polymer conformations
\cite{Hinch77}. Let us briefly repeat applicability conditions of
the equation (\ref{intro5}), discussed in \cite{01BFL}. Linearity
of the decay term in (\ref{intro5}) assumes $R\ll R_{\rm max}$,
where $R$ is the typical polymer molecule size. The equation
(\ref{intro5}) implies that there is a single mode related to the
polymer deformations, which is an idealization. A polymer molecule
has a lot of deformational degrees of freedom, that have different
relaxation times. A number of such degrees of freedom was observed
in experiments with DNA \cite{Chu1}. Nevertheless, in the turbulent
flows, only the mode with the largest relaxation time is strongly
excited, whereas other modes are excited at most weakly. Thus, Eq.
(\ref{intro5}) should be treated as the equation related to the
principal mode.

The concentration of the polymer molecules $n$ enters the system
(\ref{intro1},\ref{intro5}) only via $\Pi_0$, $\Pi_0\propto n$ and
implicitly via the assumption $R\ll R_{\rm max}$ since $\Pi\propto
nR^2$. If $n$ is inhomogeneous, then the system of equations
(\ref{intro1},\ref{intro5}) has to be supplemented by the equation
for the concentration $\partial_t n+{\bbox v}\nabla n=0$ (we
neglect small diffusivity of polymer molecules). In this paper we
consider the case when polymer molecules are strongly extended.
Then $\Pi\gg\Pi_0$, and the term with $\Pi_0$ in Eq. (\ref{intro5})
can be discarded. In this case any explicit dependence on the
concentration of the polymer molecules $n$ drops from the system of
equations (\ref{intro1},\ref{intro5}). Therefore the dynamics of
the polymer solutions with different values of $n$ is identical in
this regime as long as $n$ is large enough for the condition $R\ll
R_{\rm max}$ to be satisfied.

\subsection{Lagrangian Description}
\label{subsec:lagrange}

One can establish some properties of the polymer stress tensor $\Pi$,
using the Lagrangian description of a fluid. It is based on the
notion of fluid particles trajectories (Lagrangian trajectories)
$\bbox x(t,\bbox r)$, which are determined by the relations
\begin{eqnarray} &&
\partial_t {\bbox x}={\bbox v}(t,{\bbox x})\,,\quad
{\bbox x}(t_0,{\bbox r})={\bbox r} \,.
\label{lagrtr} \end{eqnarray}
The point ${\bbox r}$ plays the role of a Lagrangian marker. If
$\Pi_0$ in Eq. (\ref{intro5}) is neglected, then it is possible to
write its solution as
\begin{eqnarray} &&
{\Pi}(t,{\bbox x})=W(t, t_0){\Pi}(t_0,\bbox r)
W^{T}(t, t_0) e^{-{2(t-t_0)}/{\tau}} \,,
\label{intro4} \end{eqnarray}
where the superscript $T$ denotes a transposed matrix. Here $W$ is
the Lagrangian mapping matrix determined by the relations
\begin{eqnarray} &&
\partial_t W(t,t_0)=\sigma(t) W(t,t_0)\,,
\qquad W(t_0,t_0)=1\,,
\label{mmm1} \\ &&
\sigma_{ij}(t,\bbox r)
=\nabla_j v_i[t,{\bbox x}(t,\bbox r)] \,.
\label{sigma} \end{eqnarray}
Above $\sigma$ is the tensor of the velocity derivatives along the
Lagrangian trajectory $\bbox x(t)$ that includes the strain tensor
and local rotations. The incompressibility condition
$\nabla\cdot{\bbox v}=0$ is formulated in terms of $\sigma$ as
${\rm tr}\,\sigma=0$. Then a consequence of Eq. (\ref{mmm1}) is
$\mbox{det}\,W=1$.

The matrix $W$ describes deformations of infinitesimal fluid
volumes. For example, the separation, $\delta{\bbox x}$, between
two close fluid particles, moving along the Lagrangian trajectory
$\bbox x(t)$, evolves according to
\begin{eqnarray} &&
\delta{\bbox x}(t)=W(t,t')\,\delta{\bbox x}(t')\,.
\label{divl} \end{eqnarray}
Therefore $W_{ij}(t,t_0,{\bbox r})={\partial x_i(t,{\bbox
r})}/{\partial r_j}$. Now it is easy to understand a meaning of Eq.
(\ref{intro4}). The polymers are advected along the Lagrangian
trajectories being stretched by the velocity gradient and relaxing
to their equilibrium shape due to the polymer elasticity.

We now briefly describe statistical properties of the matrix $W$,
details can be found in Appendix. We represent the matrix as
$W=M\Lambda N$, where $M$ and $N$ are orthogonal matrices, while
$\Lambda$ is a diagonal matrix. At times much larger than the
velocity gradients correlation time $\tau_{\sigma}$ the main
eigenvalue $\exp(\rho_1)$ of $\Lambda$ becomes much larger than the
rest, under the condition that the set of the Lyapunov exponents
$\lambda_i$ is non-degenerate. If a statistically steady state is
realized, then the observation time is arbitrarily large, and we
conclude from Eq. (\ref{intro4}) that the matrix $\Pi$ has to be
uniaxial
\begin{eqnarray} &&
\Pi_{ik}=B_i B_k \,,
\label{bibk} \end{eqnarray}
as it was noted in \cite{01BFL}. This conclusion is almost
self-evident once one goes back to the derivation of Eq.
(\ref{intro5}), recognizing that at $R\gg R_0$ the contribution of
thermal fluctuations into $\Pi$ is negligible, so that
$\Pi_{ij}\propto R_iR_j$ holds. We observe that the vector $\bbox
B$ characterizes the direction and the strength of the coherent
molecule elongations weighted by their contribution into the stress
tensor. Note, that $\bbox B$ is defined up to sign, in analogy with
the director in nematic liquid crystals. It follows from
(\ref{intro4}) and (\ref{bibk}) that
\begin{eqnarray} &&
{\bbox B}(t,\bbox x)=\exp[-(t-t_0)/\tau]
W(t,t_0){\bbox B}(t_0,\bbox r).
\label{evol} \end{eqnarray}
If $t-t_0\gg\tau$ then $W$ in this relation can be estimated as
$e^{\rho_1}$.

There are some modifications of the $W$ statistics with respect to
a Newtonian fluid, that are imposed by the above relations. As it
follows from Eq. (\ref{evol}), stationarity of the $\bbox B$
statistics implies that $\rho_1(t)-t/\tau$ has a stationary
distribution. In particular, we conclude that the principal
Lyapunov exponent $\lambda_1=\lim_{t\to\infty}\rho_1/t$ of the flow
is equal to $1/\tau$, independently of the Reynolds number. This
means that above the coil-stretch transition the characteristic
value of the velocity gradient is fixed at the scale $1/\tau$. The
above behavior is contrasted to the Newtonian fluids for which
$\lambda_1$ grows with increasing $\mbox{Re}$ and fluctuations of
$\rho_1-t/\tau$ grow with time. The absence of the growth of the
fluctuations is related to anticorrelations in the temporal
dynamics of the component $\tilde \sigma_{11}(t)$ of $\sigma_{ij}$
along $\bbox B$. These anticorrelations show themselves in the
equality $\int\mbox dt\langle\!\langle\tilde\sigma_{11}(0)
\tilde\sigma_{11}(t)\rangle\!\rangle=0$ (double brackets designate
an irreducible correlation function) and originate in the special
interaction of the inertial and elastic degrees of freedom,
explained in more detail in Section \ref{sec:saturation}.

\subsection{Dynamic Equations for Dilute Polymer Solutions}
\label{subsec:equa}

It is convenient to rewrite the equations
(\ref{intro1},\ref{intro5}) in terms of the vector $\bbox B$ thus
getting rid of extra degrees of freedom. Substituting the
decomposition (\ref{bibk}) into Eq. (\ref{intro5}), one obtains
\begin{eqnarray} &&
\partial_t {B}_{i}+(\bbox v\nabla){B}_{i}
={B}_{k}\nabla_k v_i-B_i/\tau \,.
\label{inbb} \end{eqnarray}
This equation is similar to the one satisfied by the magnetic field
in MHD \cite{MHD}, with the constant damping instead of the
magnetic resistivity. The resemblance is made even stronger by
noting that ${\bbox B}$ has to be solenoidal. Indeed, it follows
from Eq. (\ref{inbb}) that
\begin{eqnarray} &&
\partial_t {\nabla\cdot\bbox B}
+({\bbox v}\nabla){\nabla\cdot\bbox B}
=-{\nabla\cdot\bbox B}/{\tau}\,.
\label{alf1} \end{eqnarray}
Therefore $\nabla\cdot\bbox B$ monotonically decays becoming zero
in the (statistically) steady state. Substituting the expression
(\ref{bibk}) into Eq. (\ref{intro1}), and taking into account the
constraint $\nabla\cdot\bbox B=0$, one obtains
\begin{eqnarray} &&
\partial_t {\bbox v}+({\bbox v}\nabla ){\bbox v}
=\nu \nabla^2 {\bbox v}-\varrho^{-1}\nabla P
+({\bbox B}\nabla){\bbox B}\,.
\label{ns} \end{eqnarray}
Now the analogy of the system (\ref{inbb},\ref{ns}) with the system
describing MHD \cite{MHD} at zero magnetic resistivity is almost
complete. The only difference is in the damping term in Eq.
(\ref{inbb}).

The energy density per unit mass is given by the sum of kinetic
$v^2/2$ and elastic $B^2/2$ terms. The energy balance equation,
following from Eqs. (\ref{inbb},\ref{ns}), is
\begin{eqnarray} &&
(\partial_t+\bbox v\nabla)
\left(v^2/2+B^2/2\right)=
(\bbox B\nabla)(\bbox B\cdot\bbox v)
-\varrho^{-1}\bbox v\nabla P
\nonumber \\ &&
+\nu\nabla_i(\bbox v \nabla_i\bbox v)
-\nu (\nabla_i v_k)^2 -\tau^{-1}B^2 \,.
\label{energy} \end{eqnarray}
The energy dissipation is due to the viscous and the polymer
relaxation terms. Other terms in Eq. (\ref{energy}) represent
energy fluxes (in real space), they can be written as full
divergences due to the constraints $\nabla\cdot\bbox
v=0=\nabla\cdot\bbox B$.

\section{Mechanism of scale-invariance:
passive scalar with linear damping}
\label{sec:passive}

Before investigating statistical properties of the polymer
solutions, described by the equations (\ref{inbb},\ref{ns}), we
present statistical properties of a passive scalar with constant
damping at scales smaller than the smoothness scale of the flow,
examined in \cite{98Che}. This simple case enables one to recognize
the origin of a power spectrum for passively advected fields.

The equation for the passive scalar $\theta$ in the considered case is
\begin{eqnarray} &&
\partial_t\theta+\bbox v\nabla \theta
=-\theta/\tau+\phi \,,
\label{scal1} \end{eqnarray}
where $\tau$ is the passive scalar decay time and $\phi$ is a
forcing term needed to maintain the stationary state. It is assumed
to be concentrated at a finite range of wavevectors near $k_f$. We
have omitted the diffusive term which can be neglected in
comparison with the constant damping (in some region of scales)
provided the diffusion coefficient is small enough. Note that the
constant damping of the passive scalar leads to a well-defined
steady statistics even in the absence of the passive scalar diffusion.

A flow is smooth on scales smaller, than the velocity gradients
correlation length $l$. The smoothness means that a velocity
difference between two points can be approximated by a linear profile
\begin{eqnarray} &&
\delta v_i=\sigma_{ij}\delta r_j\,,
\label{diff} \end{eqnarray}
where $\delta r_j$ is the separation between the points and
$\sigma$ is a function of time. Obviously, $\sigma_{ij}=\nabla_j
v_i$. For the usual turbulent
velocity the correlation length $l$ is equal to the viscous scale,
$l=\eta$. The smoothness of the turbulent velocity at scales less
than $\eta$ was first exploited by Batchelor \cite{Batchelor'}, who
considered statistical properties of a passive scalar at these scales.

The linearity of the velocity difference leads to a power law for
the passive scalar spectrum $E(k)$, which is defined as
$\langle\theta({\bbox k})\theta({\bbox k}')\rangle
=(2\pi)^3\delta({\bbox k}+{\bbox k}')E(k)$, where
$\theta(\bbox k)$ is Fourier transform of $\theta(\bbox r)$ with
the wave vector $\bbox k$, angular brackets designate averaging
over the statistics of ${\bbox v}$, and we assume homogeneity and 
isotropy of the statistics. Indeed, the time of the energy transfer from $k$ to
$2k$ at $kl\gg 1$ is scale-independent due to the linearity of the
velocity profile. On the other hand, during the spectral transfer
time the energy decay is also $k$-independent (since the damping
term is scale-independent). As a result the spectral function
$E(k)$ satisfies a relation $E(2k)=CE(k)$ (with a constant $C<1$).
The solution of this equation is a power-law $E(k)\propto
k^{-\alpha}$ with $2^{\alpha}C=1$.

Now we put the above consideration into a more rigorous frame. We
consider the passive scalar spectrum $E(k)$ at $kl\gg 1$. The
evolution of wave packets with such wavevectors is determined by
the velocity gradient $\sigma$. Let us consider the evolution during
a time $t_0$ and express $\theta(t_0)$ via $\theta(0)$. A value of
$\theta(t_0)$ near a point $\bbox r_1$ is determined by an
evolution of $\theta$ in the vicinity of the Lagrangian trajectory
${\bbox x}(t,{\bbox r}_1)$. To examine this evolution, one may
perform the Taylor expansion of the velocity $\bbox v$ in Eq.
(\ref{scal1}) up to the first order in $\bbox r-\bbox x$ since the
homogeneous advection does not affect equal-time correlation
functions due to the Galilean invariance. Then one obtains
\begin{eqnarray} &&
\partial_t \theta +[{\bbox u}+
\sigma\cdot({\bbox r}-{\bbox x})]\nabla\theta
=-\theta/\tau+\phi \,.
\label{fou1} \end{eqnarray}
Here $\bbox u(t)=\bbox v(t,\bbox x)$ and $\sigma=\sigma(t,\bbox x)$
are the velocity and the velocity gradients matrix along the
Lagrangian trajectory $\bbox x$. Fourier transform $\theta_k$ of
the field $\tilde\theta$ measured in the moving frame $\tilde
\theta(t,{\bbox r})\equiv\theta(t,{\bbox r}+{\bbox x})$ satisfies
\begin{eqnarray} &&
\partial_t\theta_k-\left({\bbox k}\sigma
\cdot\frac{\partial}{\partial
{\bbox k}}\right)\theta_k=-\frac{\theta_k}{\tau}
+\phi_k\exp\left[i{\bbox k}\cdot{\bbox x}\right].
\label{circum} \end{eqnarray}
Further we confine ourselves to wavevectors $k\gg k_f$, that is
much larger than those on which the pumping $\phi$ is supported. In
this case $\theta_k$ is determined by the convection from smaller
wavevectors and the forcing term can be neglected. Then the
equation (\ref{circum}) can be solved explicitly, and we find
\begin{eqnarray} &&
\theta(t, {\bbox k})=
e^{-t/\tau}\theta(0,{\bbox k}W) \,,
\label{fou0} \end{eqnarray}
where $W=W(t,0)$ is the Lagrangian mapping matrix, see Subsection
\ref{subsec:lagrange}. Returning to the real space, we obtain
\begin{eqnarray} &&
\theta(t, {\bbox r}+{\bbox x})=
e^{-t/\tau}\!\!\int\frac{\mbox d\bbox k}{(2\pi)^3}
e^{i\bbox k\left[{\bbox r}+ W{\bbox x}(0)\right]}
\theta(0,{\bbox k}W) \,,
\label{fou2} \end{eqnarray}
The above formula is valid for $|W^{-1}{\bbox
r}|\ll l$, the condition means that the passive scalar coming to a
point ${\bbox r}+{\bbox r}_1$ was all the time in the $l$-vicinity
of ${\bbox x}$ allowing the Taylor expansion for the velocity.

Let us now consider the pair correlation function of the passive
scalar $f(r)\equiv\langle \theta(t_0,{\bbox r}_1)\theta(t_0,{\bbox
r}_1+{\bbox r})\rangle$, defined as the spatial average over
${\bbox r}_1$. We assume that the average $\langle\theta\rangle$ is
zero (which can always be achieved by a shift of $\theta$ by a
constant). The product of the fields is given by Eq. (\ref{fou2})
[remind that $\bbox x(t_0)=\bbox r_1$] and depends on ${\bbox r}_1$
via the argument of $W$. The average over space (over ${\bbox
r}_1$) is equivalent to the average over the velocity statistics,
or over the velocity gradients statistics along the Lagrangian
trajectories. If $\lambda_1 t_0\gg 1$ then the average over the
interval $0<t<t_0$ and negative times can be done independently.
Indeed, $\sigma(t)$ has a Lagrangian correlation time
$\lambda_1^{-1}$. Thus velocity at negative times is correlated
with $\sigma$ only at $|t|\sim\lambda_1^{-1}$
while $W(t_0)$ is not sensitive to the value of $\sigma$ there, due
to $\lambda_1^{-1}\ll t_0$ (see Appendix). Therefore we can write
\begin{eqnarray} &&
f(r)=e^{-2t_0/\tau} \int \frac{\mbox d\bbox
k}{(2\pi)^3}\exp(i\bbox k {\bbox r})
\langle E[{\bbox k}W(t_0)]\rangle,
\label{fou9} \end{eqnarray}
where $E(k)$ is the spectrum function introduced above. Noting that
$E(k)$ equals the Fourier transform of the pair-correlation
function we obtain the following stationarity condition for the
spectrum (we substitute $t_0$ by $t$)
\begin{eqnarray} &&
E(\bbox k)=\left\langle\exp(-2t/\tau)
E({\bbox k}W)\right\rangle\,,
\label{mi3} \end{eqnarray}
where $W=W(t,0)$. The equation (\ref{mi3}) is applicable at $kl\gg
1$, as follows from the condition $r\ll l$ in Eq. (\ref{fou2}).

The relation (\ref{mi3}) has a simple meaning. The wavevectors of
small-scale fluctuations of the passive scalar evolve according to
${\bbox k}(t)={\bbox k}(0)W^{-1}(t)$ as was shown by Kraichnan
\cite{Kraichnan0}. Thus the energy of a fluctuation with the
wavevector ${\bbox k}$ is equal to its energy time $t$ ago at the
wavevector ${\bbox k}W(t,0)$ multiplied by the decaying factor
$\exp(-2t/\tau)$. Note that we could equally well start directly
from (\ref{circum}) to derive the equation (\ref{mi3}). One can
show that in the spatially homogeneous situation one can always
introduce such equation for the investigation of the spectrum at
$kl\gg 1$.

The equation (\ref{mi3}) is the quantification of the heuristic
arguments given in the beginning of the section taking into account
that the energy transfer time is by itself a random quantity. Its
solution is a power law $E(k)\propto k^{-\alpha}$. Substituting the
expression into Eq. (\ref{mi3}), one gets the relation
\begin{eqnarray} &&
\exp(2t/\tau)=\left\langle
|\bbox k W/k|^{-\alpha}\right\rangle \,,
\label{mii3} \end{eqnarray}
which determines the exponent $\alpha$. At $\lambda_1 t\gg1$ the
moments of $|\bbox k W/k|$ behave exponentially with time. Indeed,
they are roughly equal to the product of $\lambda_1 t$ independent
identically distributed random variables. Besides at these times
the moments are independent of ${\bbox k}/k$ due to the isotropization
of $W(t_0)$ described in Appendix. As a result the above
equation has a unique physical solution examined in more detail in
Appendix \ref{subsec:exp}, where the inequality $\alpha>3$ is
established. The inequality has simple meaning that the spectrum
has to decay faster than the Batchelor spectrum $k^{-3}$ ($k^{-1}$
in the spherical normalization) holding at infinite $\tau$.

Let us now extend the above results. The power law spectrum
persists, even if the relaxation time $\tau$ is $\bbox
k$-dependent, but scales as zero power of $k$, that is if $\tau$
depends on the direction of $\bbox k$ only. This dependence can be
regular (which makes the spectrum anisotropic) or random. Another
remark is that addition of an oscillating term (with $\omega_k$)
into the equation for $\theta_k$,
\begin{eqnarray} &&
\partial_t\theta_k-\left({\bbox k}\sigma\cdot
\frac{\partial}{\partial\bbox k}\right)\theta_k
=-\theta_k/\tau -i\omega_k(t)\theta_k \,,
\label{mi4} \end{eqnarray}
does not change its spectrum even though the oscillating term
breaks the scale-invariance. Indeed, let us pass from $\theta$ to
$\tilde\theta$, which is $\tilde\theta=\exp(i\varphi_k)\theta$,
with the phase, satisfying
\begin{eqnarray} &&
\partial_t\varphi_k -\left({\bbox k}\sigma\cdot
\frac{\partial}{\partial\bbox k}\right)\varphi_k =\omega_k \,.
\label{mi5} \end{eqnarray}
Then for $\tilde\theta$ we return to the equation (\ref{fou2}),
that leads to the power spectrum. It remains to note that the
spectra of $\theta$ and $\tilde \theta$ coincide. In other words,
oscillating terms conserve energy and are, consequently, irrelevant
for the energy balance.

Below we generalize the simple picture, presented in this section, to
the polymer solutions.

\section{High Reynolds Flows}
\label{sec:saturation}

Here we consider turbulence in dilute polymer solutions, when the
Reynolds number exceeds the critical value ${\rm Re}_c$,
corresponding to the coil-stretch transition. Then the polymer
molecules are strongly elongated. Two different cases are possible,
depending on the concentration of the polymer molecules $n$. If it
is very small, the elastic stresses are small in comparison with
the viscous stresses. Then the polymers are stretched to their
maximal elongation, $R_{\rm max}$, and the properties of the fluid
do not differ significantly from those of the pure solvent. Below
we consider the second, more interesting, case, when the
concentration of the polymers $n$ is large enough, so that elastic
stresses can be larger than the viscous stresses. Then the polymer
back reaction substantially modifies the flow.

Whereas in the pure solvent typical velocity gradients grow
unlimited as the Reynolds number increases, in polymer solutions
above the coil-stretch transition the balance of inertial and
elastic degrees of freedom fixes the characteristic value of the
velocity gradient at $1/\tau$. Indeed, if the instantaneous
velocity gradient exceeds $1/\tau$, it extends the polymers, so
that the elastic stress grows and damps the gradient. On the other
hand, if the velocity gradient is smaller than $1/\tau$, the
molecules contract and their influence on the flow diminishes. Then
the velocity gradients tend to grow up to the value characteristic
of the pure solvent, which is larger than $1/\tau$.
Thus the velocity gradients fluctuate near $1/\tau$, that explains
the statistically steady state realized above the coil-stretch
transition. Let us derive the condition for the existence of this
steady state, related to the existence of the maximal size $R_{\rm
max}$ of the polymer molecules. In the vicinity of the coil-stretch
transition $\nabla v\sim 1/\tau$ so that $B^2_{\rm back}\sim
\nu\nabla v\sim\nu/\tau$ as it follows from Eq. (\ref{ns}). This
leads to the condition for the existence of the back reaction
regime $B^2_{\rm max}\gg\nu/\tau$, where $B^2_{\rm max}$ is the
maximal value of the elastic stress tensor achieved at $R\sim
R_{\rm max}$. Using estimates for the microscopic parameters,
proposed in Ref. \cite{Hinch77}, one can rewrite this condition as
$n\gg(R_0R_{\rm max}^2)^{-1}$. Below we will assume that there exists
an interval in ${\rm Re}$ such that $B_{\rm max}$ exceeds the value
of $B$ prescribed by the flow. The latter increases as ${\rm Re}$ grows 
so that the condition will break down at certain ${\rm Re}$. After this
happens either polymer degradation occurs or polymers start to behave as
rigid bodies with size $R_{\rm max}$. In the latter case the fluid becomes
Newtonian again with renormalized viscosity.

We assume $V\tau\ll L$, where $V$ is the velocity at the turbulence
integral scale. Then the gradient related to the large eddies is
smaller than $\tau^{-1}$. Therefore, the large eddies do not excite
polymers, which means that the elastic stress tensor is not
correlated at these scales. Since only coherent excitations of the
elastic stress tensor can influence the flow, we conclude that the
elasticity is negligible for large eddies. The interaction of
inertial and elastic degrees of freedom becomes essential at a
scale $r_*$, where velocity gradients are of the order of $1/\tau$.
Here the energy starts to dissipate due to the polymer relaxation,
that is the inertial cascade terminates at $r\sim r_*$. Since the
velocity gradients fluctuate near the value $1/\tau$, reached at $r_*$,
at $r<r_*$ velocity difference scales linearly with the distance
that is $r_*$ is the smoothness scale of the flow. Near the
coil-stretch transition characteristic velocity gradient is
determined by the viscous scale and is of the order of $1/\tau$,
hence $r_*\sim\eta$. As the Reynolds number increases, velocity
fluctuations increase, so that the scale $r_*$ grows which is very
different from the Newtonian fluids where the smoothness scale
$\eta$ decreases with $\mbox{Re}$. As the energy input increases
the viscous energy dissipation rate, $\nu(\nabla{\bbox v})^2$,
remains of the order of $\nu/\tau^2$. Therefore far above the
transition the principal part of the energy is dissipated by the
polymer relaxation. Then the viscous term in Eq. (\ref{energy}) can
be neglected and we obtain
\begin{eqnarray} &&
\langle B^2\rangle =\epsilon\tau \,,
\label{averpi} \end{eqnarray}
where $\epsilon$ is the energy injection rate per unit mass,
estimated as $V^3/L$. The relation (\ref{averpi}) means that a
typical value of $B$ grows as the energy input increases, and,
consequently, the elastic stress tensor does.

The above quantities can be estimated, using the Kolmogorov-like
reasoning. Then we obtain from Eq. (\ref{averpi}) that $B\sim
\sqrt{\epsilon\tau}$. Next, as follows from Eq. (\ref{ns}), at the
scale $r_*$ we have $v\sim B$. Equating then the characteristic
velocity gradient $v/r_*$ to $1/\tau$, one obtains $r_*\sim
(\epsilon\tau^3)^{1/2}$. Note that this agrees with the direct Kolmogorov
theory estimate of $r_*$ based on $\delta v(r_*)\sim(\epsilon r_*)^{1/3}
\sim r_*/\tau$. Near the coil-stretch transition the
viscous and elastic dissipation terms in the energy balance
equation (\ref{energy}) are of the same order. Estimating
$\epsilon$ by the viscous dissipation term $\nu/\tau^2$ one finds
$\mbox{Re}_c\sim[L^2/(\nu\tau)]^{2/3}$ for the value
of the Reynolds number at the transition. The same answer can be
found by equating $r_*$ and the Kolmogorov-41 estimate
$(\nu^3/\epsilon)^{1/4}$ for the viscous scale $\eta$.

Below we investigate the case $\mbox{Re}\gg\mbox{Re}_c$ that
elucidates most clearly the role of the polymer elasticity. Since
the condition implies that the viscous term is negligible at the
scale $r_*$, a new interval of scales, where viscosity is
negligible but elasticity is not, has to exist below $r_*$. The
analogy with the magnetic hydrodynamics, noted above, helps us to
understand dynamics of fluctuations in this interval. These
small-scale fluctuations, which occur on the background of stresses
excited at $r\sim r_*$, are elastic waves similar to the Alfven
waves propagating in the presence of a large-scale magnetic field
in plasma \cite{MHD,Kraichnan}. The dispersion relation for the
waves is $\omega=\bbox B \bbox k$, where $\omega$ is the wave
frequency and $\bbox k$ is its wave vector. Therefore the group
velocity of these waves is $\bbox B$ which can be estimated in
accordance with Eq. (\ref{averpi}) as $\sqrt{\epsilon\tau}$. The
wave velocity fluctuates, but the fluctuations occur at times
$\sim\tau$ and are slower than the wave oscillations at $k
r_*\gg1$, showing that the waves are well-defined. There exist two
mechanisms of the elastic waves attenuation: polymer relaxation and
viscous dissipation. The first mechanism leads to the
scale-independent attenuation $\sim\tau^{-1}$, which is smaller
than the frequency, at $kr_*\gg1$. The second mechanism produces
the attenuation $\sim\nu k^2$, which is much smaller than the
frequency for $k\eta_*\ll 1$ where $\eta_*=
\nu(\epsilon\tau)^{-1/2}$. Thus the elastic waves attenuate weakly
in the interval $r_*^{-1}\lesssim k\lesssim\eta_*^{-1}$. This
interval can be called the elastic waves range.

The dynamics at scales $r\ll r_*$ is also characterized by the
stretching that takes place at a time scale $\tau$ and is slower
than the wave's oscillations. It is this dynamics that determines
the velocity spectrum at $kr_*\gg 1$, since the wave oscillations
do not influence the spectrum, like in the example presented at the
end of Section \ref{sec:passive}. As a result, we come to a power
spectrum, which is examined in the next subsection.

\subsection{Power-law Spectrum in the elastic waves range}
\label{subsec:power}

As we explained, statistical stationarity implies, that the
velocity gradients fluctuate near $1/\tau$, the value
characteristic of the scales $r\sim r_*$. Therefore velocity
gradients have to decrease with diminishing scale $r$ at $r<r_*$,
which can be formulated as $\nabla v'\ll\nabla v$. Here ${\bbox
v'}$ is the small-scale component of the velocity containing only
harmonics with wavevectors satisfying $kr_*\gg 1$. The existence of
elastic waves at these scales leads to equipartition of kinetic and
elastic energies (see \cite{Kraichnan} and the proof below) so that
$\nabla B'\ll \nabla B$ holds too. As a result the influence of
${\bbox v}'$ and ${\bbox B}'$ on motions at scales $\sim r_*$ is
negligible, that is $\bbox v'$ and $\bbox B'$ can be treated as
passively advected and stretched by $\bbox v$ and $\bbox B$.
Equations for $\bbox v'$ and $\bbox B'$ can be found by linearizing
Eqs. (\ref{inbb},\ref{ns})
\begin{eqnarray} &&
\partial_t {\bbox B'}+({\bbox v}\nabla){\bbox B'}
+({\bbox v'}\nabla){\bbox B}
=({\bbox B}\nabla){\bbox v'}
+({\bbox B'}\nabla){\bbox v}-{\bbox B'}/{\tau},
\nonumber \\ &&
\partial_t\bbox v'+({\bbox v}\nabla ){\bbox v'}
+({\bbox v'}\nabla ){\bbox v}
\label{linear} \\ &&
=-\nabla p'+({\bbox B}\nabla){\bbox B'}
+({\bbox B'}\nabla){\bbox B} +\nu\nabla^2 \bbox v',
\nonumber \end{eqnarray}
where $p'=P'/\varrho$.

The inequalities $\nabla v'\ll \nabla v$, $\nabla B'\ll \nabla B$
imply that at $r\lesssim r_*$ the differences $\delta\bbox v$ and
$\delta\bbox B$, taken at points, separated less, than $r_*$, scale
linearly with the separation according to
\begin{eqnarray} &&
\delta v_i=\sigma_{ij}\delta r_j\,, \qquad
\delta B_i=\gamma_{ij}\delta r_j\,,
\label{diff1} \end{eqnarray}
which is a generalization of Eq. (\ref{diff}). Both matrices
$\sigma_{ik}=\nabla_k v_i$ and $\gamma_{ik}=\nabla_k B_i$ have
typical values $1/\tau$ and correlation times of the order of
$\tau$. To investigate the statistics of the small-scale components
we may use the same scheme, as was developed in Section
\ref{sec:passive} for the passive scalar, expanding the equations
(\ref{linear}) near a Lagrangian trajectory, like in Eq.
(\ref{fou1}). As we explained in Section \ref{sec:passive}, the
velocity should be expanded to the first order (because of the
Galilean invariance). The need to expand ${\bbox B}$ to the first
order too follows from the fact that zeroth order term produces
Alfven waves that do not affect energy balance of the waves of the
same type as explained at the end of Section \ref{sec:passive}.
Passing to Fourier components (of the functions of the argument
$\bbox r-\bbox x$), we obtain the equations
\begin{eqnarray} &&
\partial_t\bbox v'_k-\left(\bbox k\sigma\cdot
\frac{\partial}{\partial\bbox k}\right){\bbox v'}_k
+\left(\bbox k\gamma\cdot
\frac{\partial}{\partial\bbox k}\right){\bbox B'}_k
+\sigma {\bbox v'}_k
\nonumber \\ &&
=-i{\bbox k} p'_k
+i({\bbox B}\cdot {\bbox k}){\bbox B'}_k
+\gamma {\bbox B'}_k -\nu k^2 {\bbox v'}_k,
\label{alf4} \\ &&
\partial_t {\bbox B'}_k
-\left(\bbox k\sigma\cdot
\frac{\partial}{\partial\bbox k}\right){\bbox B'}_k
+\left(\bbox k\gamma\cdot
\frac{\partial}{\partial\bbox k}\right){\bbox v'}_k
+\gamma {\bbox v'}_k
\nonumber \\ &&
=i({\bbox B}\cdot{\bbox k}){\bbox v'}_k
+\sigma {\bbox B'}_k-\frac{{\bbox B'}_k}{\tau},
\label{alf5} \end{eqnarray}
analogous to Eq. (\ref{circum}). The quantities $\sigma$, $\gamma$
and ${\bbox B}$ in Eqs. (\ref{alf4},\ref{alf5}) are measured in the
Lagrangian frame (they are functions of time and the Lagrangian
marker). Correlation functions of the fields $\bbox v'$ and $\bbox
B'$ are defined as averages over volume (or, what is the same, over
different Lagrangian trajectories), that is over a statistics of
$\sigma$, $\gamma$ and ${\bbox B}$. There is a new ingredient in
comparison with the consideration of Section \ref{sec:passive},
which is the stretching terms like $\sigma {\bbox v'}_k$. However,
these terms preserve zero scaling dimension of the time-evolution
operator and, consequently, they are not expected to destroy the
power character of the velocity spectrum $E(k)$.

From now on we neglect the viscous term in (\ref{alf4}) which is
justifiable for not too large wavevectors (a criterion is
determined below). Using the incompressibility condition, we
express the pressure $p'_k=2i[{\bbox k}\sigma{\bbox v'}-{\bbox
k}\gamma{\bbox B'}]/k^2$. The description is significantly
simplified in terms of the Elsasser variables ${\bbox
g_{\pm}}={\bbox v'}_k\pm {\bbox B'}_k$, which satisfy
\begin{eqnarray} &&
\partial_t {\bbox g_{\pm}}
-\left(\bbox k\sigma_{\pm}\cdot
\frac{\partial}{\partial\bbox k}\right)
{\bbox g_{\pm}}=\pm i({\bbox B}
\cdot{\bbox k}){\bbox g_{\pm}}-\frac{{\bbox g_{\pm}}}{2\tau}
\label{alf6} \\ &&
+\frac{\bbox k}{k}\left(\frac{\bbox k}{k}
\sigma_{\pm}{\bbox g_{\pm}}\right)
+\left(\frac{1}{2\tau}-\sigma_{\mp}\right){\bbox g_{\mp}}
+\frac{\bbox k}{k}\left(\frac{\bbox k}{k}
\sigma_{\mp}{\bbox g_{\mp}}\right),
\nonumber \end{eqnarray}
where $\sigma_{\pm}=\sigma\mp\gamma$. Of course, the equations
(\ref{alf6}) are compatible with the conditions $\bbox k\cdot \bbox
g_\pm=0$ following from the solenoidality of ${\bbox v}$ and ${\bbox
B}$. The right-hand side of (\ref{alf6}) contains the Alfven term
$i\bbox B\cdot\bbox k$, describing the wave oscillations. As we
explained at the end of Section \ref{sec:passive}, it is convenient
to eliminate the oscillations, introducing the corresponding phase
and amplitude ${\bbox g_{\pm}}={\bbox a_{\pm}}\exp[i\varphi_{\pm}]$
with $\varphi_{\pm}$ satisfying
\begin{eqnarray} &&
\partial_t \varphi_{\pm}
-\left(\bbox k\sigma_{\pm}\cdot
\frac{\partial}{\partial\bbox k}\right)
\varphi_{\pm}=\pm ({\bbox B} \cdot{\bbox k}),
\label{alf8} \\ &&
\varphi_{\pm}=\pm \int_0^t \mbox dt'
{\bbox B}(t') W^{-1, T}_{\pm}(t')
W^T_{\pm}(t){\bbox k},
\label{alf9} \end{eqnarray}
where $\partial_t W_{\pm}= \sigma_{\pm}W_{\pm},\quad W_{\pm}(0)=1$.
The equations for the amplitudes $a_{\pm}$ are
\begin{eqnarray} &&
\partial_t {\bbox a_{\pm}}-
\left(\bbox k\sigma_{\pm}\cdot
\frac{\partial}{\partial\bbox k}\right){\bbox a_{\pm}}=
\frac{\bbox k\left(\bbox k \sigma_{\pm}
{\bbox a_{\pm}}\right)}{k^2}
-\frac{{\bbox a_{\pm}}}{2\tau}
\nonumber \\ &&
+\left[\left(\frac{1}{2\tau}
-\sigma_{\mp}\right)\!{\bbox a_{\mp}}
\!+\frac{\!\bbox k\!\left(\bbox k\sigma_{\mp}
{\bbox a_{\mp}}\right)}{k^2}\right]\exp(\mp i\phi) \,,
\label{alf10} \\ &&
\phi=\int_0^t\!\!\!\mbox dt' {\bbox k}
\left[W_{+}(t)W_+^{-1}(t')\!+\!W_{-}(t)W_-^{-1}(t')
\right]{\bbox B}(t') \,.
\nonumber \end{eqnarray}
The above equations are, again, compatible with the constraints
$\bbox k\cdot\bbox a_\pm=0$.

We observe that a characteristic time of the variations of the
amplitudes is $\tau$ while the phase in the last term varies by
$2\pi$ during the characteristic time $(kB)^{-1}\ll \tau$ (the last
inequality coincides with the previously derived condition for the
existence of waves). Indeed, the exponent appearing in the last
line of Eq. (\ref{alf10}) at $t\gg\tau$ can be estimated as
$B(t)\tau k e^{t/\tau}$, where $k e^{t/\tau}$ is the current value
of the wave vector, increasing due to the stretching process.
Averaging (\ref{alf10}) over times much larger than $(kB)^{-1}$ but
much smaller than $\tau$ (the procedure is nothing but the
Bogolubov-Krylov averaging method), we find the following amplitude
equation
\begin{eqnarray} &&
\partial_t {\bbox a_{\pm}}-
\left(\bbox k\sigma_{\pm}\cdot
\frac{\partial}{\partial\bbox k}\right){\bbox a_{\pm}}=
\frac{\bbox k}{k}\left(\frac{\bbox k}{k}
\sigma_{\pm}{\bbox a_{\pm}}\right)
-\frac{{\bbox a_{\pm}}}{2\tau}.
\label{basc} \end{eqnarray}
We observe that the equations for ${\bbox a}_+$ and ${\bbox a}_-$
decouple. This is in accordance with the qualitative considerations
of Kraichnan \cite{Kraichnan} who argued that the interaction of
the waves, described by the amplitudes $a_+$ and $a_-$, is weak
because their propagation directions are reverse.

The equation (\ref{basc}) has the same structure as the equation
for the linearly decaying scalar, considered in Section
\ref{sec:passive}. The difference is in its vectorial nature and in
the presence of the term directed along ${\bbox k}$ that comes from
the solenoidality condition $\bbox k \bbox a_\pm=0$.
A formal solution of the equation (\ref{basc}) can be written as
\begin{eqnarray} &&
{\bbox a_{\pm}}(t, {\bbox k})\!\!=\!\!e^{-t/2\tau}\!
M_{\pm}\left[t, \bbox q_\pm\right]\!
{\bbox a_{\pm}}\!\left[0,\bbox q_\pm\right],
\label{alf11} \\ &&
\partial_t{M}_{\pm}(t, {\bbox k})
=(f_\pm^{-2}{\bbox f}_{\pm})
\left(\bbox f_{\pm}\sigma_{\pm}M_{\pm}\right)^T.
\label{alf12} \end{eqnarray}
where ${\bbox q}_{\pm}={\bbox k}W_{\pm}(t)$, ${\bbox
f_{\pm}}={\bbox k}W^{-1}_{\pm}(t)$, and the initial condition for
the matrices $M_\pm$ is $M_{ik}(t=0,\bbox k)=\delta_{ik}-k_i
k_k/k^2$. The term $k_ik_k/k^2$ in the initial condition for
$M_{\pm}$ vanishes after contraction with solenoidal field ${\bbox
a}(t, {\bbox k})$, that leads to the correct initial condition for
$\bbox a_\pm$. Note that $\bbox f_\pm M_\pm(t,\bbox k)=0$. Indeed,
$\bbox k M_\pm(0,\bbox k)=0$ and $\partial_t [\bbox f_\pm
M_\pm(t,\bbox k)]=0$, as follows from Eq. (\ref{alf12}) and the
equations $\partial_t \bbox f_\pm=-\bbox f_\pm\sigma_\pm$.

Remind that $\bbox B$ is defined up to sign. Therefore all the
statistical properties of the solution have to be invariant under
the transformation ${\bbox B}\to-{\bbox B}$. This transformation
interchanges $\bbox g_+$ and $\bbox g_-$. Therefore statistical
properties of $\bbox g_+$ and $\bbox g_-$ are identical.
Particularly, the spectra of ${\bbox g_{+}}$ and ${\bbox g_{-}}$
coincide, that is $\langle \bbox g_{\pm i}({\bbox k})
\bbox g_{\pm j}({\bbox k'})\rangle=(2\pi)^3\delta({\bbox k}+{\bbox k'})
(\delta_{ij}-{k_i}{k_j}/k^2)E(k)$. Thus, without any loss of
generality, one can consider ${\bbox g}_+$ solely. At calculating
the correlation function of $\bbox g_+$, entering the definition of
$E$, we can exploit an independence of $W_+(t)$ and $M_+(t)$ of
${\bbox a}_{+}(0,{\bbox k})$ that holds at $t$ much larger than the
correlation time $\tau$ of $\sigma_+$ (this is again in complete
analogy with the consideration of Section \ref{sec:passive}). Then
the stationarity condition for $E({\bbox k})$ reads
\begin{eqnarray} &&
2E(\bbox k)\!=\!\exp(-t/\tau)\left\langle
Z(t, {\bbox q}_+)E(\bbox q_+)\right\rangle\,,
\label{steq} \\ &&
Z(t, {\bbox k})\!=\!\mbox{tr}\,
{M}^T_+(t, {\bbox k}){M}_+(t,{\bbox k})
\!-\!\frac{{\bbox k}\!{M}^T_+(t,{\bbox k})\!
{M}_+(t, {\bbox k}){\bbox k}}{k^2},
\nonumber \end{eqnarray}
where we used $M(t,-{\bbox k})=M(t, {\bbox k})$. Note that $Z=2$.
Indeed, using Eq. (\ref{alf12}) and $\bbox f_\pm M_\pm(t,\bbox
k)=0$ one easily shows that the time derivative of $M^T(t, {\bbox
k})M(t, {\bbox k})$ vanishes so that $M^T(t, {\bbox k})M(t, {\bbox
k})=\delta_{ij}-k_i k_j/k^2$. and $Z(t, {\bbox k})=2$. Thus the
equation (\ref{steq}) simplifies to
\begin{eqnarray} &&
E(\bbox k)=\exp(-t/\tau)
\left\langle E(\bbox q_+)\right\rangle \,,
\label{mai3}\end{eqnarray}
almost identical to Eq. (\ref{mi3}) established in Section
\ref{sec:passive}. Similarly, one can formulate an equation for the
cross-correlation spectrum function $E'({\bbox k})$ defined by
$\langle (g_+)_i({\bbox k})(g_-)_j({\bbox k'})\rangle=\delta({\bbox
k} +{\bbox k'})(\delta_{ij}-{k_i}{k_j}/k^2)E'(k)$. The fast
oscillating phase does not cancel in this equation leading to the
inequality $E'(k)\ll E(k)$. It leads to the conclusion that the
spectra of ${\bbox v}$ and ${\bbox B}$ coincide and are equal to
$E(k)/2$ each. This proves the equipartion of the elastic and the
kinetic energies claimed above.

In analogy with the consideration of Section \ref{sec:passive} one
can establish that the solution of (\ref{mai3}) is power-like
$E(k)\propto k^{-\alpha}$ where $\alpha$ is determined implicitly by
\begin{eqnarray} &&
1=\exp(-t/\tau)\left\langle
\frac{1}{|{\bbox k}W_+(t)/k|^{\alpha}}
\right\rangle \,.
\label{alf15} \end{eqnarray}
Again, $\alpha>3$ (see Appendix \ref{subsec:exp}). In fact a
stronger inequality follows from the stationarity condition.
Namely, the spectrum has to decay faster than $k^{-5}$ ($k^{-3}$ in
the spherical normalization). Otherwise $\langle(\nabla{\bbox
v})^2\rangle=\int E(k) k^2\mbox d{\bbox k}$ is determined by scales
smaller than $r_*$, violating the condition that the gradients have
to be saturated at the value $1/\tau$ reached at $r_*$. The
condition $\alpha>5$ coincides with the applicability condition of
the above consideration that uses ${\bbox v}({\bbox x}+{\bbox
r})-{\bbox v}({\bbox x})\approx\sigma {\bbox r}$ for $r\ll r_*$.
Indeed, then $\int E(k)k^2\mbox d{\bbox k}$ is determined by
$kr_*\lesssim 1$ and $\langle( {\bbox v} ({\bbox x}+{\bbox
r})-{\bbox v}({\bbox r})-\sigma {\bbox r})^2\rangle\ll\langle
(\sigma {\bbox r})^2\rangle$ for $r\ll r_*$.

It is natural to ask, whether $\alpha$ is a universal number,
independent of Re. The above analysis shows that $\alpha$ is
determined by the statistics of $\sigma$ and $\gamma$. Within the
framework of the Kolmogorov theory the statistics is independent of
the inertial interval length and can be characterized by a single
parameter $\lambda_1=1/\tau$. We conclude that the dimensionless
quantity $\alpha$ is a universal number in the Kolmogorov theory.
In fact, due to intermittency the statistics of velocity gradients
depends on the length of the inertial interval and, consequently,
on the Reynolds number. The current understanding of intermittency
does not allow us to estimate $\alpha$ for a given $\mbox{Re}$, yet
some qualitative assertions can be formulated. Intermittency
enhances the probability of large gradients which leads to faster
transfer of energy to large wavevectors. Therefore the velocity
spectrum becomes flatter as the intermittency increases. As we
established, the length of the inertial interval decreases with the
growth of the Reynolds number for polymer solutions (since the
lower boundary of the inertial interval $r_*$ increases).
Consequently, the intermittency decreases as Re grows. We conclude
that $\alpha$ should be a monotonically increasing function of
$\mbox{Re}$.

Now we establish the region of scales where the power spectrum
exists. Its lower boundary is related to the viscous dissipation,
which grows with increasing $k$, destroying the power spectrum at
large wavenumbers. Comparing the viscous term $\nu k^2 {\bbox v'}$,
the last one in Eq. (\ref{alf4}), with, say, the stretching term
$\sigma \bbox v'$, we find, that the viscous term wins at the scale
$\sim\sqrt{\nu\tau}$. As a result the power-law terminates at
$k\sim(\nu\tau)^{-1/2}$. For larger wave vectors the velocity
spectrum diminishes faster than a power of $k$, that is the power
spectrum occurs in the interval $r_*^{-1}\ll k\ll(\nu\tau)^{-1/2}$.
Note that at $\mbox{Re}\gg\mbox{Re}_c$ we have
$\sqrt{\nu\tau}\gg\eta_*$ so that the power spectrum occurs in the
interval, where the elastic waves are well-defined.

\section{Elastic turbulence}
\label{sec:elastic}

We pass to the case of low Reynolds numbers. Then a random
(chaotic) flow can be excited due to elastic instabilities, if the
Weissenberg number $\mbox{Wi}$ is larger than unity. This is the
situation of the recently discovered ``elastic turbulence''
\cite{00GS,01GSa,01GSb}.

We investigate the case $\mbox{Re}\ll 1$ where the substantial
derivative in Eq. (\ref{ns}) can be neglected. Then the system
(\ref{inbb},\ref{ns}) becomes
\begin{eqnarray} &&
\rho^{-1}\nabla P=({\bbox B}\nabla){\bbox B}
+\nu \nabla^2 {\bbox v},\quad
\nabla\cdot{\bbox v}=0,
\label{et1} \\ &&
\partial_t {\bbox B}+({\bbox v}\nabla){\bbox B}
=({\bbox B}\nabla){\bbox v}
-\bbox B/{\tau},\quad \nabla\cdot{\bbox B}=0.
\label{et2} \end{eqnarray}
The inequality $\mbox{Re}\ll 1$ implies, that the kinetic energy of
the solution can be neglected in comparison with the elastic one.
The dissipation of the elastic energy is, however, due to both
energy dissipation mechanisms (solvent viscosity and polymer
relaxation):
\begin{eqnarray} &&
\frac{\mbox d}{\mbox dt}\int \mbox d{\bbox r}\,\frac{B^2}{2}=
-\frac{1}{\tau} \int \mbox d{\bbox r}\,\frac{B^2}{2}
-\nu \int \mbox d{\bbox r}\, (\nabla_i {v}_k)^2 \,,
\label{et6} \end{eqnarray}
as follows from Eqs. (\ref{et1},\ref{et2}) and is in accordance
with Eq. (\ref{energy}).

The system of equations (\ref{et1},\ref{et2}) has to be
complemented by the boundary conditions for the velocity, which in
the absence of polymers would lead to $\mbox{Wi}>1$ so that the
equilibrium state of polymers is unstable. The instability
eventually leads to a chaotic, statistically steady state
maintained by the non-linear dynamics of $\bbox B$, see Eq.
(\ref{et2}). Stationarity of the statistics, again, implies
$\lambda_1=1/\tau$ and stationarity of the $\rho_1-t/\tau$
statistics, as it stems from Eq. (\ref{et2}). It follows that the
velocity gradients are of the order of $1/\tau$ in the bulk.
Therefore a boundary layer has to be formed, where the velocity
gradient, exceeding $1/\tau$, at the boundary, drops to the value
$1/\tau$ in the bulk. In the boundary layer the flow is mainly
shear and the polymers are weakly stretched. The existence of the
boundary layer is observed also experimentally \cite{Sasha}.

The above picture means, that instabilities lead to velocity
fluctuations with scales determined by the size of the system and a
characteristic gradient $1/\tau$. The following estimates for the
values of these large-scale fluctuations hold
\begin{eqnarray} &&
B^2\sim \nu\sigma \sim \frac{\nu}{\tau}, \quad
v^2\sim\frac{L^2}{\tau^2}, \quad
\frac{v^2}{B^2}\sim \frac{L^2}{\nu \tau}
\sim{\rm Re}\ll 1 \,,
\label{et3} \end{eqnarray}
where $L$ is the linear system size. The estimates (\ref{et3})
imply that both dissipative terms in Eq. (\ref{et6}) are of the
same order. The correlation time of ${\bbox B}$ and ${\bbox v}$ is
determined by the typical value of the stretching and is of the
order of $\tau$. The large-scale velocity fluctuations produce
smaller scale fluctuations of ${\bbox B}$ that in turn induce
small-scale fluctuations of velocity. The velocity gradients become
smaller when the scale decreases, since the large-scale velocity
gradient is of the order of $1/\tau$ and the total velocity
gradient is fixed at this value by the stationarity condition. This
is in complete analogy with the consideration of the previous
section.

Now we are going to consider statistical properties of the
small-scale fields $\bbox v'$ and $\bbox B'$. The fields evolve
passively in the large-scale fields $\bbox v$ and $\bbox B$.
However, there is a major qualitative difference from the
high-Reynolds case which is in the role of the large-scale
component of ${\bbox B}$. In the high-Reynolds number case (see
Section \ref{sec:saturation}) the terms with the large-scale field
$\bbox B$ in the equations for the small-scale fields conserve
energy and, consequently, do not enter the equation for the
spectrum (\ref{steq}). That is why the correct description of the
dynamics required account of the small gradient $\nabla_i
\bbox B$ on the background of $\bbox B$. In the elastic
turbulence case the terms with $\bbox B$ are dissipative and,
consequently, one can forget about the gradient terms, as
subleading ones. Thus the equations for the small-scale components
of the fields, following from Eqs. (\ref{et1},\ref{et2}), are
\begin{eqnarray} &&
\nabla p'=({\bbox B}\nabla){\bbox B'}
+\nu \nabla^2 {\bbox v'},
\label{et4} \\ &&
\partial_t {\bbox B}'+({\bbox v}\nabla){\bbox B'}
=({\bbox B'}\nabla){\bbox v}
+({\bbox B}\nabla){\bbox v'}-\bbox B'/{\tau} \,.
\label{et5} \end{eqnarray}
Let us stress that the above equations assume only ${B}\gg {B}'$
(not $\nabla {B}\gg \nabla {B}'$, which is in fact wrong for the
elastic turbulence).

To analyze the above equations for the small-scale components we
use the same scheme as in the previous sections. As we explained
above there is no need to account for the spatial variation of
${\bbox B}$ so that (for the Fourier components) the equations
(\ref{et4},\ref{et5}) take the form
\begin{eqnarray} &&
{\bbox v'}=\frac{i({\bbox k}\cdot
{\bbox B})}{\nu k^2}{\bbox B}',\quad p'=0,
\label{ett} \\ &&
\partial_t {\bbox B}'\!-\!\left({\bbox k}\sigma\cdot
\frac{\partial}{\partial {\bbox k}}\right){\bbox B}'
=\sigma {\bbox B}'\!-\!\frac{{\bbox B}'}{\tau}\!-\!
\frac{({\bbox k}\cdot {\bbox B})^2}{\nu k^2}{\bbox B}'.
\label{et8} \end{eqnarray}
The constraints ${\bbox k}\cdot{\bbox B}'=0$, ${\bbox k}\cdot{\bbox
v}'=0$ that follow from $\nabla\cdot\bbox v=0=\nabla\cdot\bbox B$
are consistent with the equations (\ref{ett},\ref{et8}). The last
term in the equation for ${\bbox B}'$ is due to the viscous
dissipation, the term is of the same order as the elastic
dissipation term $-{\bbox B}'/\tau$ as follows from (\ref{et3}). We
observe that, again, the time-evolution operator for ${\bbox B}'$
has zero scaling dimension and (in accordance with the discussion
at the end of Section \ref{sec:passive}) one expects that the
spectrum $F(k)$ of ${\bbox B}$ obeys a power-law. To demonstrate
this power behavior we use a formal solution of the equation (\ref{et8})
\begin{eqnarray} &&
{\bbox B}'(t,\bbox k)=W{\bbox B}'(0, \bbox k W)
\exp\left[-\frac{t}{\tau}-\int_0^t
\!\!\!\mbox d t' \xi(t')\right],
\label{et1e} \\ &&
\nu\xi(t')={[{\bbox B}(t'){\bbox n}(t', t, {\bbox k})]^2},\ \
\partial_{t'}{\bbox n}=-{\bbox n}\sigma
+{\bbox n}({\bbox n}\sigma{\bbox n}),
\nonumber \end{eqnarray}
where $W=W(t,0)$, and $\bbox n(t',t,\bbox k)$ is determined by the
above equation with the final condition ${\bbox n}_k(t,t,{\bbox
k})={\bbox k}/k$.

Let us analyze the stress spectrum function $F(k)$: $\langle
B'_i({\bbox k})B'_j({\bbox k}')\rangle=\delta({\bbox k} +{\bbox
k}')F(k)(\delta_{ij}-k_ik_j/k^2)$. Assuming that $t$ is much larger
than the correlation time $\tau$ of ${\bbox B}$ and $\sigma$, we
may average independently over the velocity statistics at negative
and positive times, as we did in the previous sections. Then we find
\begin{eqnarray} &&
2F(\bbox k)=\langle Z
\exp\left[-2\int_0^t\mbox dt'\,\xi(t')\right]
F(\bbox k W) \rangle\,,
\label{ee1} \\ &&
Z=\exp(-2t/\tau)\left[\mbox{tr} W^TW
-\frac{\bbox k[WW^T]^2\bbox k}
{\bbox k WW^T\bbox k}\right].
\label{zzz} \end{eqnarray}
The coefficient of $F$ in the right-hand side of Eq. (\ref{ee1}) is
independent of the absolute value $k$ of $\bbox k$. Therefore Eq.
(\ref{ee1}) admits a power solution $F(k)\propto k^{-\beta}$, where
$\beta$ has to be determined from the equation
\begin{eqnarray} &&
2=\left\langle \frac{Z
\exp\left[-2\int_0^t\mbox d t'\,\xi(t')\right]}
{|\bbox kW(t)/k|^{\beta}}\right\rangle.
\label{beta} \end{eqnarray}
Note that at $t\gg\tau$ the average in Eq. (\ref{beta}) is determined
by the events with $\bbox k$ directed along the eigen vector of the
matrix $WW^T$, corresponding to the smallest eigen value (in this
case the denominator in the expression achieves a minimum). Then
the second term in the right-hand side of Eq. (\ref{zzz}) can be
neglected, and we obtain $Z\approx\exp(2\rho_1-2t/\tau)$
(therefore, due to the stationarity of the $\rho_1-t/\tau$
probability distribution, a statistics of $Z$ is time-independent).
The positive noise $\xi(t)$ can be considered as stationary with
the correlation time $\tau$. Indeed, it follows from Eq.
(\ref{et1e}) that ${\bbox n}(t')$ forgets its final direction (and
thus also becomes ${\bbox k}$-independent) at $t-t'\gg\tau$. Thus
the situation is similar to the one analyzed in Sections
\ref{sec:passive},\ref{subsec:power}. Again, we can prove the
inequality $\beta>3$, see Appendix \ref{subsec:exp}.

We are now in a position to establish the spectrum of the velocity
$E(k)$. Indeed, it follows from Eq. (\ref{ett}) that $E(k)\propto
k^{-\beta-2}$ and the spherically normalized spectra obey
\begin{eqnarray} &&
E_{sph}(k)\sim v^2L (kL)^{-\beta}, \
F_{sph}(k)\sim B^2L (kL)^{2-\beta}\,.
\label{et11} \end{eqnarray}
The estimates for the values $v^2$ and $B^2$ are written in Eq.
(\ref{et3}). We observe that our scheme that assumes $\nabla v'\ll
\nabla {v}$ and ${B}'\ll {B}$ is self-consistent due to the
inequality $\beta>3$ proved above. It agrees with the experiment,
where $\beta$ is $3.3\div 3.5$ \cite{00GS,01GSa}.

Finally, let us discuss the question concerning the validity region
of the power spectrum. Probably, it is determined by the finite
diffusivity $\kappa$ of the polymer molecules which is described by
adding the term $\kappa\nabla^2\Pi_{ij}$ to the right-hand side of
Eq. (\ref{intro5}). Comparing this term with, say, the relaxation
term with $\tau$, one concludes that the power spectrum terminates
at $k_{\rm dif}\sim \sqrt{\kappa\tau}$. At smaller scales the
velocity spectrum diminishes much faster due to diffusivity.

\section{Conclusion}
\label{conclusion}

We have investigated properties of turbulence in dilute polymer
solutions in the cases where polymer molecules are strongly
stretched. We established power-law distributions of kinetic and
elastic energies over scales in some regions, where these
power-laws are not related to an energy or other conserved quantity
cascade (in contrast to the usual turbulence). In fact, excitation
of elastic degrees of freedom at any scale leads to energy
dissipation since the elastic dissipation is scale-independent.
However, precisely this scale-independence can lead to a
scale-ivariance in the dissipative intervals, where the flow can be
treated as smooth. Small-scale fluctuations are relatively weak and
evolve passively in the smooth flow. As a result, the evolution of
fine-scale fluctuations depends trivially on the scale and
power-law spectra are formed. Let us now describe the cases where
the above general ideas are applicable.

The first case, we examined, is the high Reynolds number flow above
the coil-stretch transition, when elastic degrees of freedom are
activated. Strong interaction between the elasticity and the flow
modifies the latter below the scale $r_*$ (at this scale the
velocity gradients are of the order of $1/\tau$), which is the new
energy dissipation scale. This scale is of the order of the
Kolmogorov scale at the transition and becomes larger as Re is
increased. At $r\gtrsim r_*$ the properties of turbulence are the
same as in Newtonian fluids. The energy cascades downscales from
the pumping scale and dissipates due to polymer relaxation at
$r\sim r_*$. The flow is smooth at $r\lesssim r_*$ with the
principal Lyapunov exponent $\lambda_1$ fixed at the value $1/\tau$
by the elastic back reaction. The fluctuations in the interval of
scales $\nu(\epsilon\tau)^{-1/2}\lesssim r\lesssim r_*$ are elastic
waves. This leads to the equipartition of the kinetic and of the
elastic energies, that is the velocity spectrum $E(k)$ and the
elastic spectrum $F(k)$ coincide at these scales. The smoothness of
the flow at $r\lesssim r_*$ leads to the conclusion that these
spectra are power-like and in the spherical normalization decrease
faster than $k^{-3}$ at $kr_*\gtrsim 1$. The power spectra
terminates at the scale $(\nu\tau)^{1/2}$, where the viscous
dissipation overcomes stretching.

The second case, we examined, is the elastic turbulence regime
\cite{00GS,01GSa,01GSb}. It is a chaotic state which is realized at
small Reynolds numbers Re. The velocity gradient imposed on the
system by the boundary conditions exceeds $1/\tau$ which activates
polymer degrees of freedom leading to hydrodynamic instability and
chaotization. Again, the power spectra $F(k)$ and $E(k)$ are
power-like in this case. However there are no elastic waves that
would lead to equipartition. The main energy is carried by the
polymers: $E(k)\sim\mbox{Re}\,(kL)^{-2} F(k)$, where $L$ is the
size of the system. The velocity spectrum $E(k)$ decays faster than
$k^{-3}$ in the spherical normalization, which corresponds to the
experimental data \cite{00GS,01GSa,01GSb}.

The above-described mechanism of forming power-law spectrum for
small-scale fluctuations in a chaotic flow seems rather general to
be realized for other systems. We expect it to occur in certain
regimes in magnetohydrodynamics, flows in liquid crystals and
low-dimensional flows on a substrate.

\acknowledgements

We thank E. Balkovsky, M. Chertkov, G. Falkovich, A. Groisman,
I. Kolokolov, and V. Steinberg for valuable discussions. A. F.
was supported by the grants of ISF and Minerva foundations.

\appendix

\section{Long-Time Lagrangian Statistics}
\label{long}

Let us briefly review the long-time statistical properties of the
Lagrangian mapping matrix $W$, determined by Eqs.
(\ref{mmm1},\ref{sigma}). We consider $W(t_+, t_-)$ at $t_+>t_-$
and assume that $t_+-t_-$ is much larger than the Lagrangian
correlation time $\tau_{\sigma}$ of the velocity derivatives matrix
(\ref{sigma}), where one expects a universal statistics \cite{Pope}.
If the velocity statistics is homogeneous in time, the probability
distribution of $W(t_+,t_-)$ depends on the difference $t_+-t_-$
only. Equation (\ref{mmm1}) implies that at $t_+-t_-\gg\tau_\sigma$
the matrix $W$ is a product of a large number of independent
matrices. This is the main reason for the universality of the $W$
statistics.

It is convenient to decompose the matrix $W$ as
\begin{eqnarray} &&
W(t_+, t_-)=M\Lambda N \,,
\label{product} \end{eqnarray}
where $\Lambda$ is a diagonal matrix, and $M$ and $N$ are
orthogonal matrices \cite{Orszag}. We denote the diagonal elements
of $\Lambda$ as $e^{\rho_1}$, $e^{\rho_2}$, and $e^{\rho_3}$, and
assume that they are ordered: $\rho_1>\rho_2>\rho_3$. As a
consequence of the constraint $\mbox{det}\, W=1$ we have
$\rho_1+\rho_2+\rho_3=0$. Equation (\ref{mmm1}) can be rewritten in
terms of $\rho_i$, and the matrices $M$ and $N$. The equations for
$\rho_i$ are
\begin{eqnarray} &&
{\partial\rho_i}/{\partial t_+}=\tilde\sigma_{ii}\,,
\label{rho} \end{eqnarray}
where $\tilde\sigma=M^T\sigma M$ and no summation over the
repeating index $i$ is implied. The matrices $M$ and $N$ satisfy
$\partial_t N=\Omega_1 N$ and $\partial_t M=M\Omega_2$, where
\begin{eqnarray} &&
(\Omega_1)_{ik}=\frac{\tilde\sigma_{ik}+\tilde\sigma_{ki}}
{2\sinh(\rho_i-\rho_k)}\,,\quad
(\Omega_2)_{ik}=\frac{\tilde\sigma_{ik}e^{2\rho_k}
+\tilde\sigma_{ki}e^{2\rho_i}}
{e^{2\rho_k}-e^{2\rho_i}}\,,
\nonumber \end{eqnarray}
for $i\not= k$ and $(\Omega_1)_{ik}=(\Omega_2)_{ik}=0$ for $i=k$.
It is possible to show that the eigenvalues of $W$ repel each
other, so that the inequalities $e^{\rho_1}\gg e^{\rho_2}\gg
e^{\rho_3}$ are satisfied at $t_+-t_-\gg\tau_\sigma$ \cite{99BF}.
Then the matrix $\Omega_1$ tends to zero exponentially fast, i.e.
$N$ is determined by times of the order of $\tau_{\sigma}$ in the
vicinity of $t_-$. The matrix $\Omega_2$ becomes $\rho$-independent
at $t_+-t_-\gg\tau_\sigma$ and the evolution of $M$ is decoupled
from that of $\rho_i$. Then the value of $M$ is determined by the
time of the order of $\tau_\sigma$ at $t\approx t_+$, i.e. at
$t_+-t_-\gg\tau_\sigma$ it becomes $t_-$-independent.

The solution of Eq. (\ref{rho}) is
\begin{equation}
\rho_i=\int_{t_-}^{t_+}\mbox dt'\,
\tilde\sigma_{ii}(t') \,.
\label{sol} \end{equation}
where the right-hand side of Eq. (\ref{sol}) is an integral of a
random process independent of $\rho_i$. Equation (\ref{sol}) shows
that the variables $\rho_i$ fluctuate around their average values
$\lambda_i(t_+-t_-)$. Here the constants $\lambda_i$ are equal to
$\left\langle\tilde\sigma_{ii}\right\rangle$. They are called the
Lyapunov exponents of the flow. Generally, the spectrum of the
Lyapunov exponents is non-degenerate:
$\lambda_1>\lambda_2>\lambda_3$, which is a necessary condition for
the formalism to be self-consistent. The incompressibility
condition ensures the identity $\lambda_1+\lambda_2+\lambda_3=0$,
which implies $\lambda_1>0$ and $\lambda_3<0$. Using the relation
(\ref{divl}) one can show that $\lambda_1$ is indeed the average
logarithmic divergence rate of two nearby Lagrangian trajectories:
\begin{eqnarray} &&
\left\langle{\mbox d\ln|\delta{\bbox x}|}
/{\mbox dt}\right\rangle =\lambda_1 \,.
\nonumber \end{eqnarray}
Similarly, $\lambda_1+\lambda_2=-\lambda_3$ is the average logarithmic
rate of the area growth.

Note that at $t_+-t_-\gg \tau_{\sigma}$ the statistics of $M$,
$\Lambda$ and $N$ are independent. Indeed, the values of $\rho_i$
are accumulated during the whole evolution time $t_+-t_-$ (see
(\ref{sol})) and are not sensitive both to the interval $(t_-,
t_-+\tau_{\sigma})$ determining $N$ and interval
$(t_+-\tau_{\sigma}, t_+)$ determining $M$. Both matrices $M$ and
$N$ are distributed isotropically because of the assumed isotropy
of the velocity statistics (this is the isotropization of $W$ referred
in the main text).

\subsection{Lagrangian Statistics in usual
Turbulent or Random Flows}
\label{subsec:usual}

Here we consider the Lagrangian statistics in the case of a usual
random (turbulent or chaotic) flow, when the velocity has finite
correlation time and no constraints are imposed on the flow. Then
the quantity $\rho_i$ can be treated as a sum of a large number of
independent random variables, provided $t_+-t_-\gg\tau_\sigma$. It
is known from the statistical mechanics (see, e.g., \cite{Landau})
that the distribution of such quantities is given by the exponent
of an extensive function. In our case the probability distribution
function (PDF) of $\rho_i$ is
\begin{eqnarray} &&
{\cal P}(t, \rho_1, \rho_2,
\rho_3)\propto \frac{1}{t}
\exp\left[-tS\left(\frac{\rho_1-\lambda_1t}{t},
\frac{\rho_3-\lambda_3 t}{t}\right)\right]
\nonumber \\ && \times
\delta(\rho_1+\rho_2+\rho_3)\,,
\label{pdf1} \end{eqnarray}
where $t=t_+-t_-$ and $\rho_1>\rho_2>\rho_3$ is implied
\cite{99BF}. The main exponential factor of the PDF has a
self-similar form described by the function $S$, which can be
called entropy function (see \cite{99BF,CFKV99,Ellis}). It is
positive, convex and has a minimum at zero values of its arguments.
The precise form of $S$ is determined by details of the velocity
statistics. The PDF has a sharp maximum at $\rho_i=\lambda_i t$. In
its vicinity the function $S$ has a quadratic expansion, i.e. the
distribution of $\rho$ is Gaussian. However, if one is interested
in the expectation values of exponential functions of $\rho_i$,
they are determined by the wings of the PDF where the Gaussian
approximation is invalid. This entails the use of the whole entropy
function.

To average the functions of $\rho_1$ only, one can introduce the
reduced probability distribution function,
\begin{eqnarray}&&
{\cal P}(t,\rho_1)\propto
\frac{1}{\sqrt{t}}\exp
\left[-tS_1\left(\frac{\rho_1
-\lambda_1 t}{t}\right)\right] \,,
\label{dis1} \end{eqnarray}
which is an integral of ${\cal P}(t,\rho_1,\rho_2,\rho_3)$ over
$\rho_2$ and $\rho_3$. At small $x$ the function $S_1(x)$ can be
written as
\begin{eqnarray} &&
S_1(x)\approx {x^2}/(2\Delta) \,.
\label{parab} \end{eqnarray}
Here $\Delta=\int\!\mbox dt\,\langle\!\langle\tilde
\sigma_{11}(t)\tilde\sigma_{11}(0)\rangle\!\rangle$ (where double
brackets designate irreducible correlation function) determines the
dispersion of $\rho_1$: $\langle(\rho_1-\lambda_1
t)^2\rangle\approx t\Delta$. Expansion (\ref{parab}) is sufficient
to describe typical fluctuations of $\rho_1$, whereas the whole
function $S$ is needed to describe rare events.

\subsection{Special properties of the long-time Lagrangian statistics
above the coil-stretch transition}
\label{subsec:special}

Above the coil-stretch transition the Lagrangian statistics
acquires new qualitative features caused by the polymer back
reaction. They can be inferred from the equation (\ref{evol})
which leads to
\begin{eqnarray} &&
\ln |B(t, {\bbox x})|-\ln
|B(t_0, {\bbox r})|\approx
\rho_1(t, t_0)-\frac{t-t_0}{\tau}.
\label{bb1} \end{eqnarray}
The expression (\ref{bb1}) is correct, provided $t-t_0\gg\tau$
(since the polymer relaxation time $\tau$ determines also the
velocity gradients correlation time). The left-hand side of the
equation (\ref{bb1}) has stationary statistics which leads to
dramatic changes in the statistics of $\rho_1$. Upon averaging one
finds $\langle \rho_1\rangle=t/\tau$ which means
$\lambda_1=\langle\tilde\sigma_{11}\rangle=1/\tau$ as it was
already explained in the main text. Stationarity of the dispersion
of $\rho_1$ leads to the conclusion that
$\int_{-t/2}^{t/2}\mbox dt'\langle\!\langle\tilde\sigma_{11}(0)
\tilde\sigma_{11}(t')\rangle\!\rangle\propto t^{-1}$
in the limit $t\to \infty$. This is related to the anticorrelation
property of $\tilde\sigma_{11}$, mentioned in the main text. That
means that the dispersion $\Delta$, defined by Eq. (\ref{parab}),
vanishes above the coil-stretch transition. In fact, this vanishing
is not abrupt and occurs within $1/\ln(R_{\rm back}/R_0)$ vicinity
of ${\mbox Re}_c$. Finally one concludes that
$\langle\!\langle\tilde\sigma_{11}(0)\tilde\sigma_{11}(t)
\rangle\!\rangle\propto t^{-2}$ at large $t$ which is again very different
from a Newtonian fluid where exponential decay is observed, see
Ref. \cite{Pope}.

\subsection{Inequality for the exponents $\alpha$, $\beta$}
\label{subsec:exp}

Here we establish the inequality for the exponent $\alpha$,
characterizing the passive scalar spectrum, see Section
\ref{sec:passive}, and the inequality for the exponent $\beta$
appearing in the elastic turbulence problem, see Section
\ref{sec:elastic}. In both cases we investigate the solution
$\Delta$ of the equation of the type
\begin{eqnarray} &&
\left\langle |{\bbox k}W|^{\Delta}
\exp\left[-\int_0^t\mbox d t' y(t')\right]
\right\rangle\sim k^{\Delta},
\label{Delta} \end{eqnarray}
where $t$ is much larger than the correlation time of a random positive
noise $y(t)$ and $\sigma$. At these times the behavior of the
moments is exponential and one can define
$\langle \exp[-\int_0^t\mbox d t' y(t')]|{\bbox k}W|^{\delta}\rangle
\sim k^{\delta}\exp[\tilde \gamma(\delta)t]$ where $\tilde\gamma(\delta)$
is a convex function due to H\"older inequality. This function is
strictly smaller than another convex function $\gamma(\delta)$
defined by $\langle |{\bbox k}W|^{\delta}\rangle \sim
k^{\delta}\exp[\gamma(\delta)t]$.

The function $\gamma(\delta)$ has a universal behavior which we
describe now. It is convenient to write
\begin{eqnarray} &&
\left\langle|{\bbox k}W|^{\delta}\right\rangle
=\int\mbox d{\bbox k}' k'^{\delta}
\left\langle \delta\left({\bbox k}-{\bbox k}'W^{-1}(t)\right)
\right\rangle\,,
\label{mi6} \end{eqnarray}
making it explicit that the wavevectors evolve according to ${\bbox
k}(t)={\bbox k}(0)W^{-1}(t)$ \cite{Kraichnan0}. Introducing ${\bbox
k}'W^{-1}(t)=k'\exp\left[\rho(t)\right]{\bbox n}$, where ${\bbox n}$
is a unit vector, one finds
\begin{eqnarray} &&
\rho(t)=\int_0^t\mbox dt' \zeta(t'),\quad
\frac{\mbox d{\bbox n}}{\mbox dt}
={\bbox n}\sigma+{\bbox n}\zeta,\ \
\zeta\equiv {\bbox n}\sigma{\bbox n}.
\nonumber \end{eqnarray}
One observes that $\zeta$ is independent of $\rho(0)$ which leads
to the conclusion that at $t\gg \tau_c(\sigma)$, where
$\tau_c(\sigma)$ is the correlation time of $\sigma$, the
probability distribution of $\rho$ is described by an entropy
function $S$.

It can be shown that $\langle\rho(t)\rangle=|\lambda_3| t$, where
$|\lambda_3|$ is the lowest in the hierarchy of the Lyapunov
exponents of the flow. This fact is intuitively clear as
$\lambda_3$ determines contraction in the real space and thus
stretching in $k$-space. We note that $\rho(t)$ is determined by
the whole interval $(0, t)$ while ${\bbox n}(t)$ only by
$\lambda_1^{-1}$ vicinity of $t$. As a result, $\rho$ and ${\bbox
n}$ are independent at $\lambda_1 t\gg 1$. Since ${\bbox n}$ is
isotropically distributed over the unit sphere one finds
$\langle\delta[{\bbox k}-{\bbox k}'W^{-1}(t)]\rangle=\langle
\delta(\rho-\ln(k/k'))\exp[-3\rho]\rangle/(4\pi k'^3)$. Substituting
this into (\ref{mi6}) and performing the integral one finds
\begin{eqnarray} &&
\int\! \frac{\mbox d\rho}{N}\exp\left[-(\delta+3)\rho
-tS\left(\frac{\rho-|\lambda_3|t}{t}\right)\right]
\sim \exp[\gamma(\delta)t],
\nonumber \end{eqnarray}
where $N\propto \sqrt{t}$ is the normalization factor insignificant
for the following considerations. It follows that besides the zero
at the origin following from the definition, the function
$\gamma(\delta)$ vanishes at $\delta=-3$. This is a general
consequence of the isotropy employed above, see also
\cite{Zeldovich}. Besides, we observe that $\gamma'(-3)=
\lambda_3<0$. Combining these properties with the convexity of
$\gamma(\delta)$ we conclude that $\gamma(\delta)$ is negative for
$-3<\delta<0$ and positive otherwise.

To prove the inequality on $\Delta$ appearing in Eq. (\ref{Delta})
one notes that both $\gamma(\delta)$ and $\tilde \gamma(\delta)$
tend to $\infty$ as $|\delta|\to \infty$. Then it follows from
$\tilde \gamma(\delta)<\gamma(\delta)$ that there are two solutions
of Eq. (\ref{Delta}): one positive and one smaller than $-3$.
Substituting $y(t)=2/\tau$ and recognizing that the exponent
$\alpha$ appearing in (\ref{mii3}) must be positive we conclude
that $\alpha>3$. Analogously, substituting $y(t)=2\xi(t)$ we
conclude that the solution of Eq. (\ref{beta}) satisfies $\beta>3$.

Finally, let us give an example of calculating $\alpha$ in a limiting
case. The exponent is determined by the equation
\begin{eqnarray}&&
\int\!\! \frac{\mbox d\rho}{N}\!\exp\left[(\alpha-3)\rho-tS\left(
\frac{\rho-|\lambda_3|t}{t}\right)\right]\!=\!\exp\left[\!
\frac{2t}{\tau}\!\right].\!
\label{mi7}\end{eqnarray}
Note that $\alpha-3$ vanishes in the limit $|\lambda_3|\tau
\to\infty$ since in this limit the linear decay term is negligible
in Eq. (\ref{scal1}) and Batchelor $k^{-3}$ spectrum must result.
Therefore at large $|\lambda_3|\tau$ the integral (\ref{mi7}) is
determined by the maximum of the probability concentrated at
$\rho=|\lambda_3|t$ which leads to
\begin{eqnarray}&&
E(k)\sim k^{-3-2(|\lambda_3|\tau)^{-1}},\ |\lambda_3|\tau\gg 1.
\label{mi8} \end{eqnarray}
For a general value of $|\lambda_3|\tau$ the exponent $\alpha$ is
determined by the concrete form of the entropy function.

\end{multicols}

\end{document}